\documentclass[fleqn,usenatbib]{mnras}

\usepackage{ae,aecompl}

\usepackage{graphicx}	
\usepackage{amsmath}	
\usepackage{amssymb}	

\newcommand{\msun}{M_\odot}

\title[Planets In Young Massive Clusters]{On the survivability of planets in young massive clusters and its implication of planet orbital architectures in globular clusters}

\author[M. X. Cai et al.]{
Maxwell X. Cai$^{1}$\thanks{E-mail: cai@strw.leidenuniv.nl (MXC)},
S. Portegies Zwart$^{1}$,
M.B.N. Kouwenhoven$^{2}$,
Rainer Spurzem$^{3,4,5}$
\\
$^{1}$Leiden Observatory, Leiden University, PO Box 9513, 2300 RA, Leiden, The Netherlands\\
$^{2}$Department of Mathematical Sciences, Xi{'}an Jiaotong-Liverpool University, 111 Ren{'}ai Rd., Suzhou Dushu Lake Science \\and Education Innovation District, Suzhou Industrial Park, Suzhou 215123, P.R. China\\
$^{3}$National Astronomical Observatories and Key Laboratory of Computational Astrophysics, Chinese Academy of Sciences, 20A Datun Road, Chaoyang District, Beijing 100012, P.R. China\\
$^{4}$Kavli Institute for Astronomy and Astrophysics, Peking University, 5 Yi He Yuan Road, Haidian District, Beijing 100871, P.R. China\\
$^{5}$Zentrum f\"{u}r Astronomie, Astronomisches Rechen-Institut, University of Heidelberg, M\"{o}nchhofstrasse 12-14, D-69120 Heidelberg, Germany\\
}

\date{Accepted XXX. Received YYY; in original form ZZZ}

\pubyear{2019}

\begin{document}
\label{firstpage}
\pagerange{\pageref{firstpage}--\pageref{lastpage}}
\maketitle

\begin{abstract}
As of August 2019, among the more than 4000 confirmed exoplanets, only one has been detected in a globular cluster (GC) M4. The scarce of exoplanet detections motivates us to employ direct $N$-body simulations to investigate the dynamical stability of planets in young massive clusters (YMCs), which are potentially the progenitors of GCs. In an $N=128{\rm k}$ cluster of virial radius 1.7 pc (comparable to Westerlund-1), our simulations show that most wide-orbit planets ($a\geq 20$~au) will be ejected within a timescale of 10 Myr. Interestingly, more than $70\%$ of planets with $a<5$~au survive in the 100 Myr simulations. Ignoring planet-planet scattering and tidal damping, the survivability at $t$ Myr as a function of initial semi-major axis $a_0$ in au in such a YMC can be described as $f_{\rm surv}(a_0, t)=-0.33 \log_{10}(a_0) \left(1 - e^{-0.0482t} \right) + 1$. Upon ejection, about $28.8\%$ of free-floating planets (FFPs) have sufficient speeds to escape from the host cluster at a crossing timescale. The other FFPs will remain bound to the cluster potential, but the subsequent dynamical evolution of the stellar system can result in the delayed ejection of FFPs from the host cluster. Although a full investigation of planet population in GCs requires extending the simulations to multi-Gyr, our results suggest that wide-orbit planets and free-floating planets are unlikely to be found in GCs.
\end{abstract}

\begin{keywords}
methods: numerical -- planets and satellites: dynamical evolution and stability -- planets and satellites: formation -- galaxies: star clusters: general -- globular clusters: general.
\end{keywords}


\section{Introduction}
\label{sec:intro}
\begin{figure}
	\centering
	\includegraphics[scale=0.5]{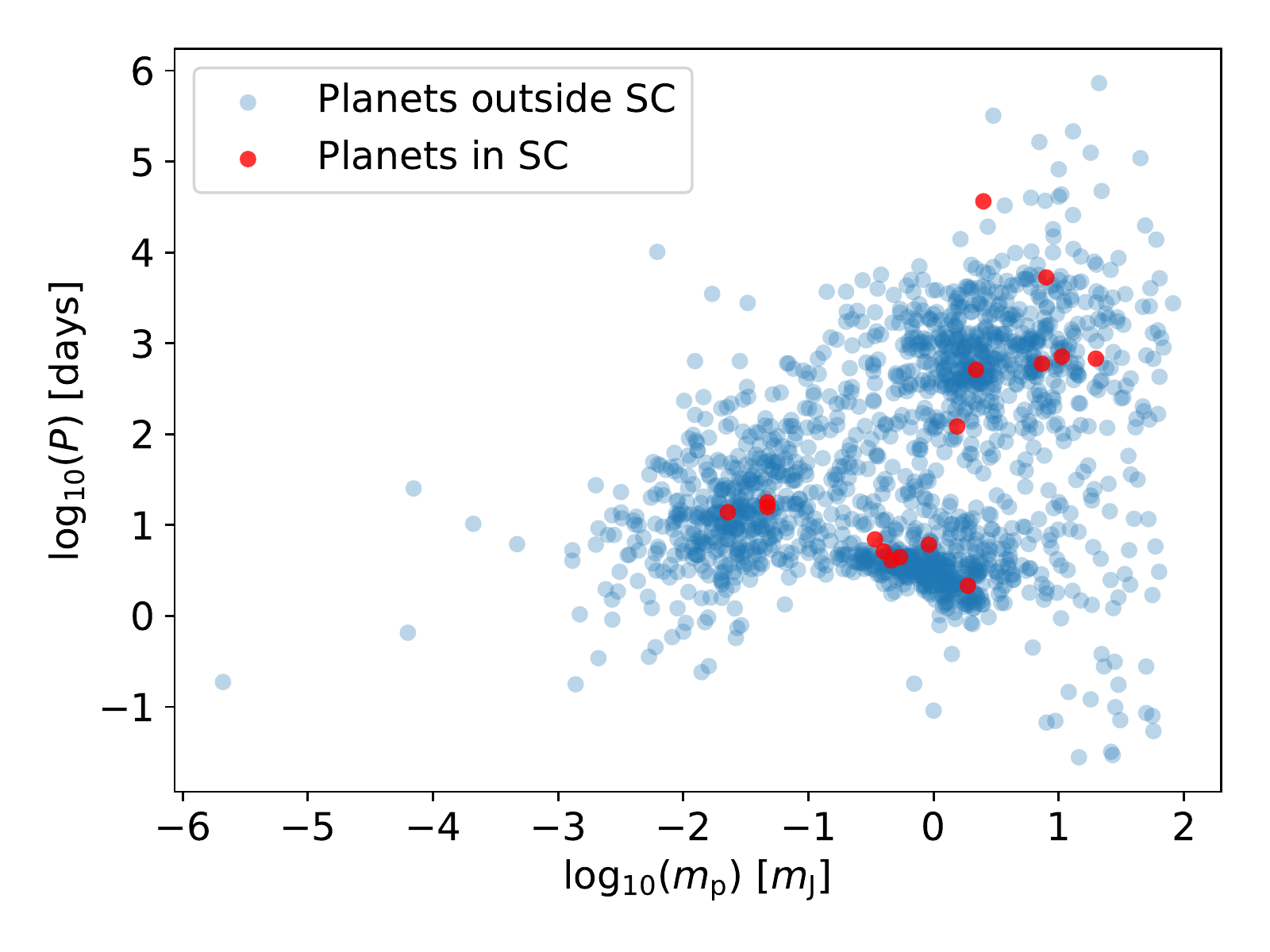}
	\caption{The masses (in Jupiter mass $M_{\rm J}$) and orbital periods (in days) of planets inside/outside star clusters (error bars not shown). Only confirmed planets with well-determined masses and orbital periods are shown. Exoplanet data downloaded from \url{exoplanet.eu} on 7 August 2019.}
	\label{fig:p_in_out_sc}
\end{figure}

 It is now widely accepted that planets are prevalent in the universe \citep{2015ARA&A..53..409W}. The planet formation process is typically considered as a byproduct of the star formation process \citep[e.g.,][]{2010A&A...510A..72F}, which takes place mainly in star clusters and stellar associations \citep[e.g.,][]{2003ARA&A..41...57L}. However, as of July 2019, with more than 4,000 confirmed exoplanets listed in the exoplanet encyclopedia\footnote{\url{http://exoplanet.eu}}, only 30 of which are detected in star cluster (see Table~\ref{tab:p_sys_sc}), and PSR B1620-26 (AB) b is the only one detected in the dense globular clusters (GCs). The exoplanets detected in star clusters are plotted in Fig.~\ref{fig:p_in_out_sc} against exoplanets detected outside star clusters. Apart from the difference in numbers, planets detected inside and outside star clusters seems to be statistically indistinguishable, with the exception that the uppermost point to the right representing PSR B1620-26 b (see Table~\ref{tab:p_sys_sc} for its orbital parameters) is probably formed by a dynamical interaction \citep{2000ApJ...528..336F,2003Sci...301..193S} in the dense core of Messier 4. The low number of planet detection in star clusters seems to be contradictory to the theoretical expectation \citep{2003ARA&A..41...57L}. The search of planets in dense star clusters started in from the late 1990s, when researchers use the Hubble Space Telescope to observe the 47 Tucanae cluster for 8.3 days. No planet is detected despite that the cluster has 33,000 stars \citep{2000ApJ...545L..47G,2017AJ....153..187M}. It is certainly possible that observational biases are responsible, but various analysis on planets in star clusters \citep[e.g.,][]{2006ApJ...641..504A,2007MNRAS.378.1207M,2009ApJS..185..486P,2009ApJ...697..458S,2011MNRAS.411..859M,2012MNRAS.419.2448P,2013MNRAS.433..867H,2015MNRAS.448..344L,2017MNRAS.470.4337C,2018MNRAS.474.5114C, 2019arXiv190204652V} have demonstrated that the dense star cluster environments (in particular, globular clusters) have implications to the formation and evolution of planets.

\begin{table*}
	\caption{List of exoplanet detections in star clusters, sorted chronologically according to the years of detection. DM: detection method; TS: transit; RV: radial velocity; TM: timing; Nep: Neptune-sized; $M_{\rm S}$: Stellar mass in solar units $\msun$; $m_{\rm p}$: planet mass in Jupiter units $M_{\rm J}$; $P$: orbital period in days; TBC: to be confirmed; *: the system K2-136 is a binary system, and therefore there are two host-star mass components. $a$: orbital semi-major axis of the planet in au; $e$: orbital eccentricity of the planet; $i$: inclination of the planet. }
	\begin{tabular}{lrrrrrrlll}
		\hline
		\hline
		{\bf Designation} & {\bf $m_{\rm p}$ ($M_{\rm J}$)} & {\bf $P$ (days)} & {\bf $M_{\rm S}$ ($\msun$)} & $a$ {\bf (au)} & $e$ & $i$ [deg] & {\bf DM} &  {\bf Cluster} & {\bf Ref.$^{\ddag}$} \\
		\hline
		\hline
		K2-264 b          & sub-Nep       & 5.84   & 0.471          & 0.05 &  0   & 88.9 & TS & Praesepe (M44)     & [1][2] \\
		K2-264 c          & sub-Nep       & 19.66  & 0.471          & 0.11 &  0   & 89.6 & TS & Praesepe (M44)     & [1][2] \\
		K2-231 b          & 0.0227        & 13.84  & 1.01           & --   & --   & 88.6 & TS & Ruprecht 147       & [3] \\	
		K2-136 b          & super-Earth   & 7.98   & $0.74/0.1^{*}$ & --   & 0.1  & 89.3 & TS & Hyades             & [4] \\
		K2-136 c          & super-Earth   & 17.3   & $0.74/0.1^{*}$ & --   & 0.13 & 89.6 & TS & Hyades             & [4][5] \\
		K2-136 d          & super-Earth   & 25.58  & $0.74/0.1^{*}$ & --   & 0.14 & 89.4 & TS & Hyades             & [4] \\	
		SAND978 b         & 2.18          & 511.2  & 1.37           & --   & --   & --   & RV & M67                & [6] \\
		EPIC 211913977 b  & sub-Nep       & 14.68  & 0.8            & --   & 0.1  & 89.4 & TS & Praesepe (M44)     & [7] \\
		EPIC 211970147 b  & super-Earth   & 9.92   & 0.77           & --   & 0.1  & --   & TS & Praesepe (M44)     & [8] \\
		YBP401 b          & 0.46          & 4.09   & 1.14           & --   & 0.15 & --   & RV & M67                & [9] \\	
		Pr 0211 c         & 7.95          & 5\,300 & 0.935          & 5.8  & 0.7  & --   & RV & Praesepe (M44)     & [10] \\
		K2-95 b           & 1.67          & 10.13  & 0.43           & 0.065& 0.16 & 89.3 & TS & Praesepe (M44)     & [8][20] \\
		EPIC 211969807 b  & sub-Nep       & 1.97   & 0.51           & --   & 0.18 & 88   & TS & Praesepe (M44)     & [8] \\
		EPIC 211822797 b  & sub-Nep       & 21.17  & 0.61           & --   & 0.18 & 89.5 & TS & Praesepe (M44)     & [8] \\
		K2-100 b          & sub-Nep       & 1.67   & 1.18           & --   & 0.24 & 85.1 & TS & Praesepe (M44)     & [8] \\
		K2-77 b           & 1.9           & 8.2    & 0.8            & --   & 0.14 & 88.7 & TS & Praesepe (M44)     & [11] \\
		EPIC 211901114 b (TBC)   & $<5$   & 1.64   & 0.46           & --   & --   & --   & TS & Praesepe (M44)     & [8] \\
		EPIC 210490365 b  & $< 3$         & 3.48   & 0.29           & --   & 0.27 & 88.3 & TS & Hyades             & [12]  \\
		SAND 364 b        & 1.54          & 121.7  & 1.35           & --   & 0.35 & --   & RV & M67                & [13]  \\
		YBP1194 b         & 0.34          & 6.96   & 1.01           & --   & 0.24 & --   & RV & M67                & [13] \\
		YBP1514 b         & 0.4           & 5.11   & 0.96           & --   & 0.39 & --   & RV & M67                & [13] \\
		HD 285507 b       & 0.92          & 6.08   & 0.73           & 0.073& 0.086& --   & RV & Hyades             & [14]\\
		Kepler-66 b       & 0.047         & 17.82  & 1.04           & 0.135& --   & --   & TS & NGC6811            & [15] \\
		Kepler-67 b       & 0.047         & 15.73  & 0.87           & 0.12 & --   & --   & TS & NGC6811            & [15] \\
		Pr 0201 b         & 0.54          & 4.33   & 1.234          & --   & --   & --   & RV & Praesepe (M44)     & [16] \\
		Pr 0211 b         & 1.88          & 2.15   & 0.935          & 0.03 & 0.017& --   & RV & Praesepe (M44)     & [16]  \\
		eps Tau b         & 7.34          & 595    & 2.70           & 1.9  & 0.15 & --   & RV & Hyades             & [17]  \\
		NGC 2423 3 b\dag      & 10.6          & 714    & 2.40           & 2.1  & 0.21 & --   & RV & NGC2423            & [18][21] \\
		NGC 4349 127 b\dag    & 19.8          & 678    & 3.90           & 2.38 & 0.19 & --   & RV & NGC4349            & [18][21] \\
		PSR B1620-26 (AB) b    & 2.50     & 36\,525 & 1.35          & 23   & --   & --   & TM & M4                 & [19] \\
		\hline
		\hline
	\end{tabular}
	\begin{flushleft}
		\dag: See also a recent analysis by \cite{2018A&A...619A...2D}, which suggests a non-planet origin NGC 4349 No. 127 and NGC 2423 No. 3.\\
		$^{\ddag}$ References: [1] \cite{2018AJ....156..195R}; [2] \cite{2019MNRAS.484....8L}; [3] \cite{2018curtis}; [4] \cite{2018mann}; [5] \cite{2018ciardi}; [6] \cite{2017brucalassi}; [7] \cite{2017mann}; [8] \cite{2017mann2}; [9] \cite{2016Brucalassi}; [10] \cite{2016malavolta}; [11] \cite{2017MNRAS.464..850G}; [12] \cite{2016mann}; [13] \cite{2014Brucalassi}; [14] \cite{2014Quinn}; [15] \cite{2013Meibom}; [16] \cite{2012Quinn}; [17] \cite{2007Sato}; [18] \cite{2007Lovis}; [19] \cite{1993Backer}; [20] \cite{2016AJ....152..223O}; [21] \cite{2018A&A...619A...2D}. 
	\end{flushleft}
	
	\label{tab:p_sys_sc}
\end{table*}

\begin{figure*}
	\centering
	\includegraphics[scale=0.5]{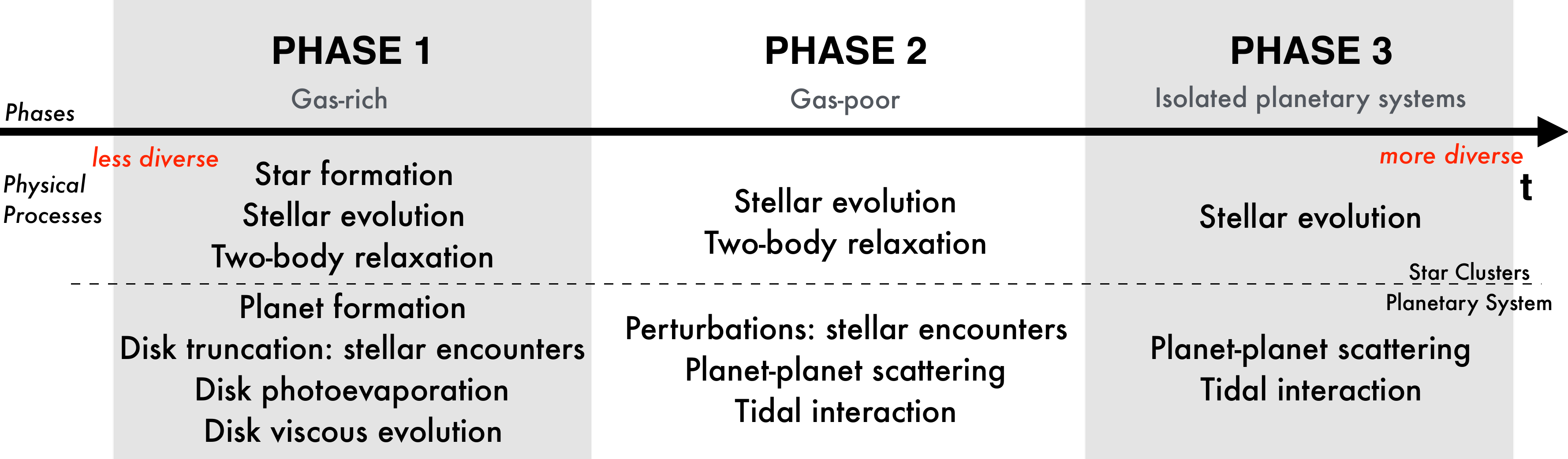}
	\caption{The coevolution of planetary systems in (massive) star clusters can be roughly divided into three phases. The diversity of exoplanets emerges gradually as a function of time.}
	\label{fig:physical_processes}
\end{figure*}

One could roughly divide the entire formation and evolution history of planets in star clusters, as shown in Fig~\ref{fig:physical_processes}. Planetary systems may be influenced by their birth environments in the following ways: first, during the planet formation process (Phase 1), protoplanetary discs may be photoevaporated due to the possible presence of nearby OB stars \citep[e.g.,][]{1999ApJ...515..669S,2000A&A...362..968A,2010ARA&A..48...47A,2013ApJ...774....9A,2016MNRAS.457.3593F} and/or truncated due to stellar encounters \citep[e.g.,][]{1993MNRAS.261..190C,1994ApJ...424..292O,2006ApJ...642.1140O,2016MNRAS.457..313P,2019MNRAS.482..732C}, which in turn modifies the properties of the discs and ultimately causes different outcome of planet formation; second: after the planet formation process (Phase 2), planets are no longer protected by the damping of the gaseous discs, and their orbital inclinations and eccentricities can be excited or even ejected by stellar flybys. Planetary systems in the dense regions of the cluster have lower chances to survive \citep{2009ApJ...697..458S, 2013MNRAS.433..867H, 2015MNRAS.448..344L, 2017MNRAS.470.4337C, 2019arXiv190204652V,2019arXiv190807747F}. Moreover, surviving planets from high-density regions tend to have relatively high mean eccentricities and inclinations due to their stellar encounter histories \citep{2018MNRAS.474.5114C}. Planet-planet scattering \citep[e.g.,][]{2009ApJ...696L..98R} and secular resonance \citep[e.g.,][]{2001ApJ...558..392R} continue long after the cluster dissolves or the planetary system leaves the cluster (Phase 3). The field exoplanets observed nowadays are the surviving planets in this natural selection process; their diversity is therefore in part shaped by their diverse birth environments in the parental cluster.

Therefore, to take into account the complex processes that eventually lead to the observable population of planets in the present day, one would have to start the simulations when planets are evolving with the host clusters in the early stage. In this study, we want to derive the orbital architectures of exoplanets in the present-day GCs, so we instead simulate the dynamical evolution of planets in young massive clusters (YMCs), which are considered as the progenitors of GCs \citep[e.g.,][]{2014CQGra..31x4006K}. YMCs are dense stellar systems of $\ge 10^4 M_{\odot}$, and with a typical age of just a few Myr \citep{2010ARA&A..48..431P,2014prpl.conf..291L}. Due to their young ages, a significant fraction of gas content is still presented inside the cluster, and the star formation process is likely to be ongoing.  

On the other hand, star formation is typically followed by planet formation \citep[e.g.,][]{2010A&A...510A..72F}. It is now widely accepted that planets are prevalent in the universe \citep{2015ARA&A..53..409W}. If planets did form in an ancient population of YMCs, and that they indeed have evolved into GCs that we observed in the present-day, will the planets that formed at the YMC stage be inherited by the GC? In another word, will the present-day GCs be a place to hunt for planets? If so, what would be the orbital architecture that we could expect?

The primary objective of this paper is to study the dynamical stability and orbital architectures of planetary systems in dense YMCs comparable to Westerlund-1, and the dynamical signature of free-floating planets (FFPs) in YMCs. This paper is organized as follows: the modeling approach and the initial conditions are presented in Section~\ref{sec:modeling_ic}, the results are presented in Section~\ref{sec:results}, followed by discussions in Section~\ref{sec:discussions}. Finally, the main conclusions are summarized in Section~\ref{sec:conclusions}.

\section{Initial Conditions and Simulations}
\label{sec:modeling_ic}
Numerical simulations of planetary systems in YMCs are constrained by three factors: First, planetary systems are chaotic few-body systems, and therefore we need to carry out a grid of simulations that covers a certain parameter space and obtain the results statistically, rather than drawing conclusions from a single simulation; Second, direct $N$-body simulations with stellar evolution taken into account are generally required to simulate the collisional dynamics of YMCs, especially if the host YMC is rotating, having sub-structures, and/or diverting from virial equilibrium; Third, there is a hierarchical timestepping problem when evolving planetary systems in YMCs: Due to the very different dynamical timescale (days or years in planetary systems and millions of years in star clusters), the integrator is forced to adapt very small time step to resolve the planetary systems accurately, which is prohibitively expensive for integrating the host clusters.

We develop a GPU-accelerated hybrid code to tackle these challenges. We realize that star clusters and planetary systems are very different, and that they have to be modeled with their own dedicated algorithms. We first integrate the host YMCs (without planets) using \texttt{NBODY6++GPU} \citep{1999JCoAM.109..407S,2003gnbs.book.....A,2015MNRAS.450.4070W}. The simulation is stored at a very high time resolution using an incremental adaptive storage scheme \citep{2012NewA...17..520F,2015ApJS..219...31C}. The storage scheme, namely ``Block time step (BTS) storage scheme'', stores only the most recently updated particles, and thereby allows very high temporal resolution with reasonable file sizes. With the BTS data, we are then able to reconstruct the details of close encounters; the close encounters details are then inserted into the \texttt{IAS15} integrator \citep{2015MNRAS.446.1424R} of the planetary system dynamics code \texttt{rebound} \citep{2012A&A...537A.128R}. The planetary system integrator queries the position vector of the closest neighbor at a timestep of years. Such a query is implemented by interpolating the BTS data on the GPUs (graphics processing units). The communication of perturbation data is implemented with the \texttt{AMUSE}\footnote{\url{https://github.com/amusecode/amuse}} \citep{2013A&A...557A..84P,2013CoPhC.184..456P,2018araa.book.....P} framework. Given that the density of YMCs is very high, especially in the cluster center, we include 5 nearest perturbers\footnote{We have performed a convergence test on the number of perturbers. We find that for a YMC like Westerlund-1, the result converges with the inclusion of 5 or more perturbers. Including 5 perturbers provides a reasonable tradeoff of accuracy and speed.}.


For simplicity, the initial condition of the star cluster is sampled from an $N=128{\rm k}$ Plummer model \citep{1911MNRAS..71..460P} in virial equilibrium (i.e. $Q=0.5$) with a \cite{2001MNRAS.322..231K} initial mass function. We consider the mass range of stars from $0.08M_{\rm \odot}$ to $100M_{\rm \odot}$. The mean stellar mass is $0.58M_{\odot}$. We assume $10\%$ of primordial binaries, but assume that all planets are orbiting single stars. The binary population has a thermal eccentricity distribution. The logarithmic of semi-major axes of binaries distribute uniformly from in the range of [$3.5 \times 10^{-8}, 3.5 \times 10^{-4}$]~parsec (0.007-72 au). The two member-stars of a binary system is equal-mass. The cluster is subject to the Galactic tidal field in the Solar neighborhood. The star cluster has an initial virial radius of $R_{\rm vir} = 1.74$ parsec, which is comparable to Westerlund-1 \citep{2010ARA&A..48..431P}. A more realistic set of YMC initial conditions that include substructures, gas, and non-virial equilibrium states will be addressed in an upcoming study.


The diversity of exoplanets clearly shows that there is no ``typical'' architecture for planetary systems. A planetary system may have multiple super-Earths or mini-Neptune packed in densely populated orbits \citep[resembles the Kepler-11 system, see][]{2011Natur.470...53L}, or it may have a lot of nearly massless objects spreading over a range of semi-major axis (e.g., the Kuiper belt). Inspired by these two distinctively different systems, we construct the following two models of idealized planetary systems:
\begin{itemize}
	\item {\bf Model-A:} Massless multi-planet system: the system consists of 50 test particles. The semi-major axes of these planets range from 6 au to 400 au. All planets are initially on circular and coplanar orbits. In this model, external perturbations due to stellar flybys play an exclusive role in shaping the orbits of the test particles.   
	\item {\bf Model-B:} Equal-mass multi-planet system: the system consists of 5 equal-mass planets, each of which is $3M_{\oplus}$, separated with 15 mutual Hill Radii with the adjacent planet. The innermost planet has an initial semi-major axis of 0.5 au, and the outermost planet has an initial semi-major axis of 6 au. All planets are initially on circular and coplanar orbits. The system is stable without external perturbations, but its stability can be hampered when the planets are excited to higher eccentricities due to stellar flybys. In this model, we would like to study the combined effect of external perturbation and internal planet-planet scattering.
\end{itemize}

We create an ensemble of 200 identical realizations of each model and place them randomly around Solar-type stars (i.e., $1 M_{\odot}$) in the host cluster, which results in 400 simulations. Each simulation is carried out for a timescale of 100 Myr. The simulations are carried out automatically using the Simulation Monitoring tool \texttt{SiMon} \citep{2017PASP..129i4503Q}.

\section{Results}
\label{sec:results}
\subsection{Surviving Orbital Architectures}
\begin{figure}
	\centering
	\includegraphics[scale=0.55]{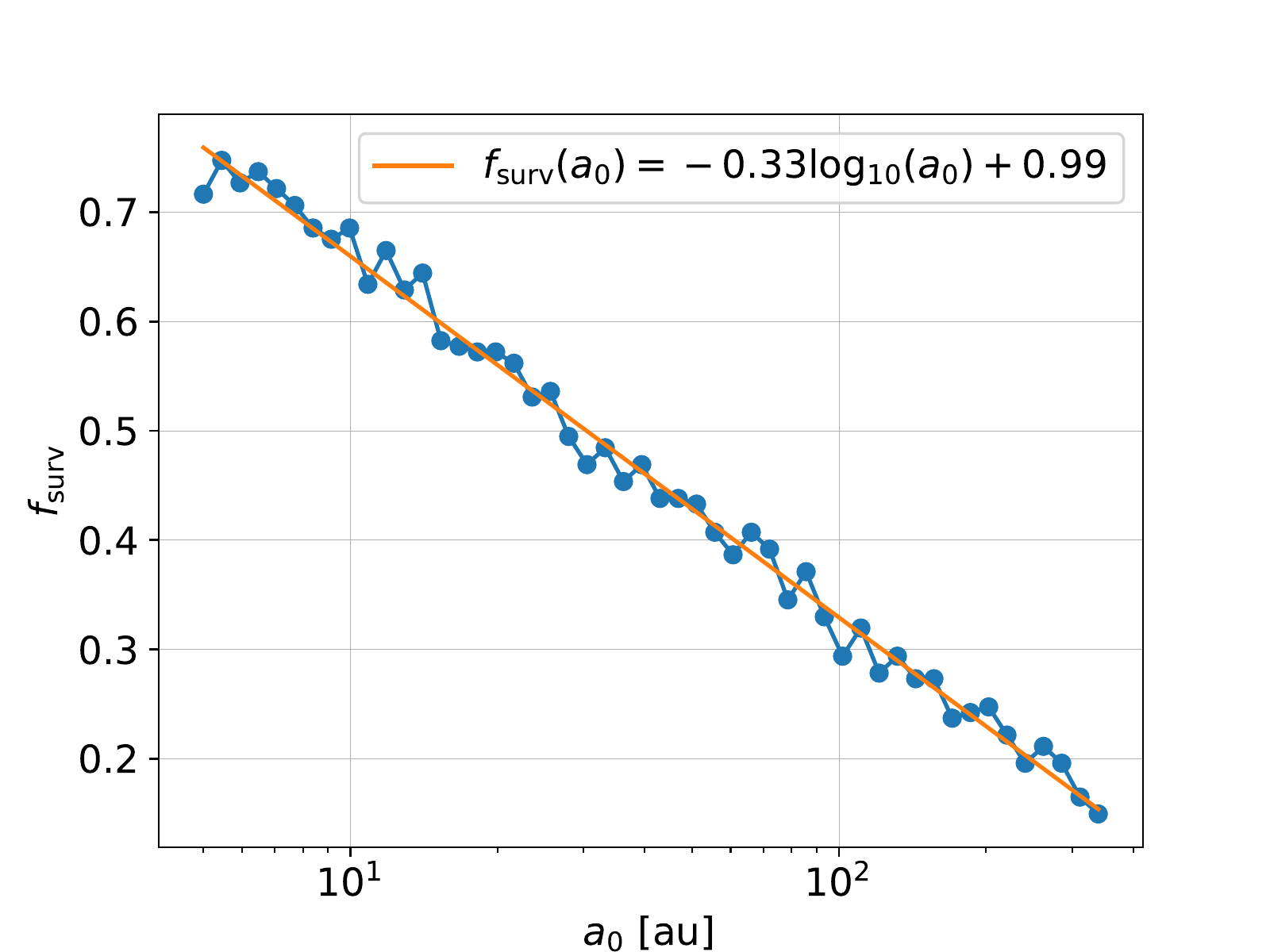}
	\caption{Survival probability of planets in YMCs as a function of the initial semi-major axes (for Model-A). The simulation results are shown with dots, and the fitting is shown with a line.}
	\label{fig:f_surv_a}
\end{figure}

We obtain statistical results from the simulation ensembles. The evolution of planetary systems in star clusters can be considered as a natural selection process, in the sense that only those planetary systems with suitable orbital architectures will be able to survive until the end of the simulations. Since Model-A surveys a wide range of semi-major axes from 6-350 au, we are able to obtain a relatively smooth distribution of survival rates as a function of initial semi-major axes, as shown in Fig.~\ref{fig:f_surv_a}. As intuitively expected, exoplanets with larger initial semi-major axes are more prone to excitation and ejections. The survival rate decreases sharply with the increment of the semi-major axis. The profile of surviving rates as a function of $\log_{10}(a_0)$ can be fitted with a linear decay function. The best fit yields:
\begin{equation}
	f_{\rm surv}(a_0) = -0.33 \log_{10}(a_0) + 1.
	\label{eq:f_surv_a0}
\end{equation}
Here, $a_0$ is in the unit of au. The fitting above is valid for $a \geq 1$~au. For planets with $a<1$~au, we expect the \emph{direct} effect of the stellar environments to be mostly negligible (note, however, that if planet-planet interaction is not negligible, a short-period planet with $a_0 < 1$ can be indirectly affected by stellar encounters when an outer planet through close encounters with the inner planet; see the results of Model-B systems). Ignoring planet-planet scattering, for a planet with a semi-major axis of $\sim 30$~au (comparable to Neptune's semi-major axis), the probability for it to be ejected is more than 0.5; the probability for a wide planet of $a>300$~au to survive in a YMC is nearly zero.  We expect the survival rates to be lower in among multi-planet systems where planet-planet interactions are important \citep[cf.][]{2017MNRAS.470.4337C}. In this sense, the exoplanets listed in Table~\ref{tab:p_sys_sc}, most of which being short-period planets, are not entirely due to observational biases, because long-period planets indeed have difficulties to survive in dense clusters. However, the survivability can be enhanced by tidal damping at the perihelion of a highly eccentricity orbit, causing the planet to significantly shrink its orbital semi-major axis and become a short-period planet \citep{2016ApJ...816...59S}.

\begin{figure}
	\centering
	\includegraphics[scale=0.6]{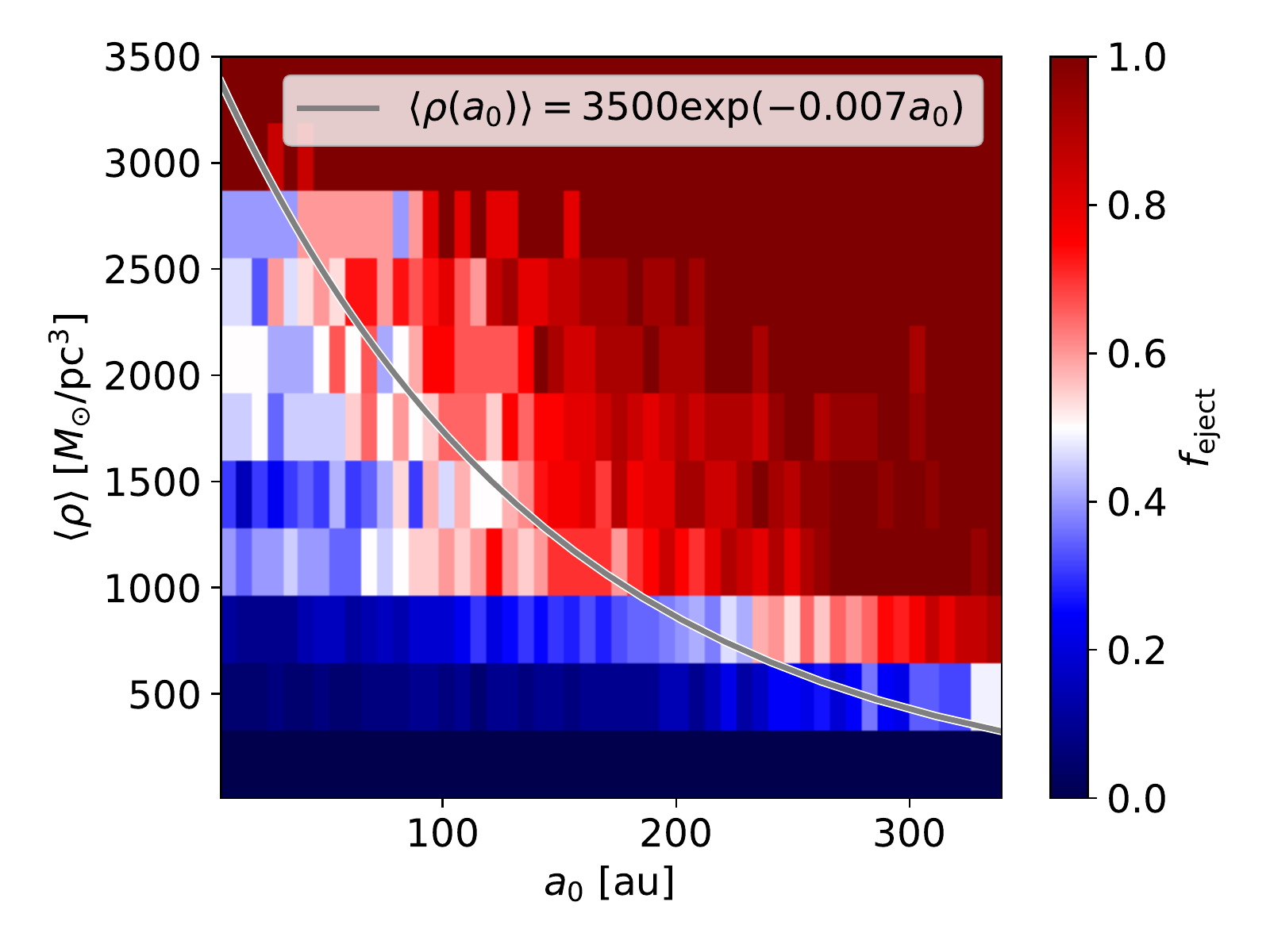}
	\caption{Survivability matrix of planets (Model A) in YMCs as a function of the initial semi-major axes $a_0$  and the mean stellar density $\bar{\rho}$ in the vicinity of the host star. Note that in our simulations we had only a few planetary systems that experienced the extremes in mean density $\bar{\rho}$ and as a result the very top as well as the bottom of the figure are affected by small-numer statistics. The grey curve roughly correspond to the $\langle \rho \rangle - a_0$ relation in which the ejection rate is constant (i.e. $f_{\rm eject} = 0.5$).}
	\label{fig:surv_matrix}
\end{figure}

Recently, \cite{2015MNRAS.451..144P} and \cite{2018MNRAS.474.5114C} suggest that the orbital architecture of surviving planetary systems (in the field) can be used to constraint its birth environments. Inspired by this idea, in Fig.~\ref{fig:surv_matrix} we plot the survival rate matrix as a function of both initial semi-major axis and the mean stellar density in the vicinity of the planetary system. The stellar mass density of the host YMC is defined according to the Plummer density profile \citep[e.g.,][]{1987degc.book.....S}:
\begin{equation}
	\rho(r_i) = \left( \frac{3 M}{4\pi a_{\rm P}^3} \right) \left( 1 + \frac{r_i^2}{a_{\rm P}^2} \right)^{-5/2},
	\label{eq:rho_r_i}
\end{equation}
where $r_i$ is the distance between the perturbed planetary system and the cluster center, $M$ is the total mass of the cluster, and $a_{\rm P}$ is the Plummer scale length. A time series of $r_i, i=1,2,3,...$ is obtained through the YMC simulation using \texttt{NBODY6++GPU}, and the corresponding $\rho(r_i)$ is calculated according to Eq.\ref{eq:rho_r_i}. The mean stellar density along the trajectory of a planetary system during the simulation time is defined as: \footnote{Note that $\langle \rho(r_i) \rangle$ is a time-averaged density that a planetary system experiences along its trajectory during the simulation time, rather than the actual density of the host cluster. The density in the cluster center can be much higher than $\langle \rho(r_i) \rangle$.}
\begin{equation}
	\langle \rho \rangle \equiv \frac{1}{N_t} \sum_{i=1}^{N_t} \rho(r_i),
\end{equation}
where $N_t$ is the total number of time steps performed during the entire simulation. A fixed timestep of 1000 yr (comparable to $\sim 10^{-3}$ crossing time of the cluster) per snapshot is used to sample $\langle \rho \rangle$. In the limit of high mean stellar densities (at the top of the figure), the planetary system spends most of its time in the dense cluster center, and the system suffers from almost total ejection regardless of the initial semi-major axis. Planetary systems slightly outside the dense regions are allowed to survive provided that the initial semi-major axis is sufficiently small. In the outskirts of the cluster, even wide planets have fair chances to survive. However, we notice that the number of planetary systems in very high or very low mean stellar densities are small, and therefore we advise the readers not to extrapolate the data in this two regimes with low-number statistics. Interestingly, the constant $f_{\rm eject}$ curve, plotted with a grey line in Fig.~\ref{fig:surv_matrix}, can be roughly approximated with an exponential decay function. In the case of $f_{\rm eject} = 0.5$, which means half of the planets are ejected, the $\langle \rho \rangle - a_0$ relation is roughly
\begin{equation}
	\langle \rho \rangle (a_0) = 3500 \exp(-0.007a_0).
\end{equation}
Here, $\langle \rho \rangle$ is in the units of $M_{\odot}/{\rm pc}^3$.

\begin{figure}
	\centering
	\includegraphics[scale=0.4]{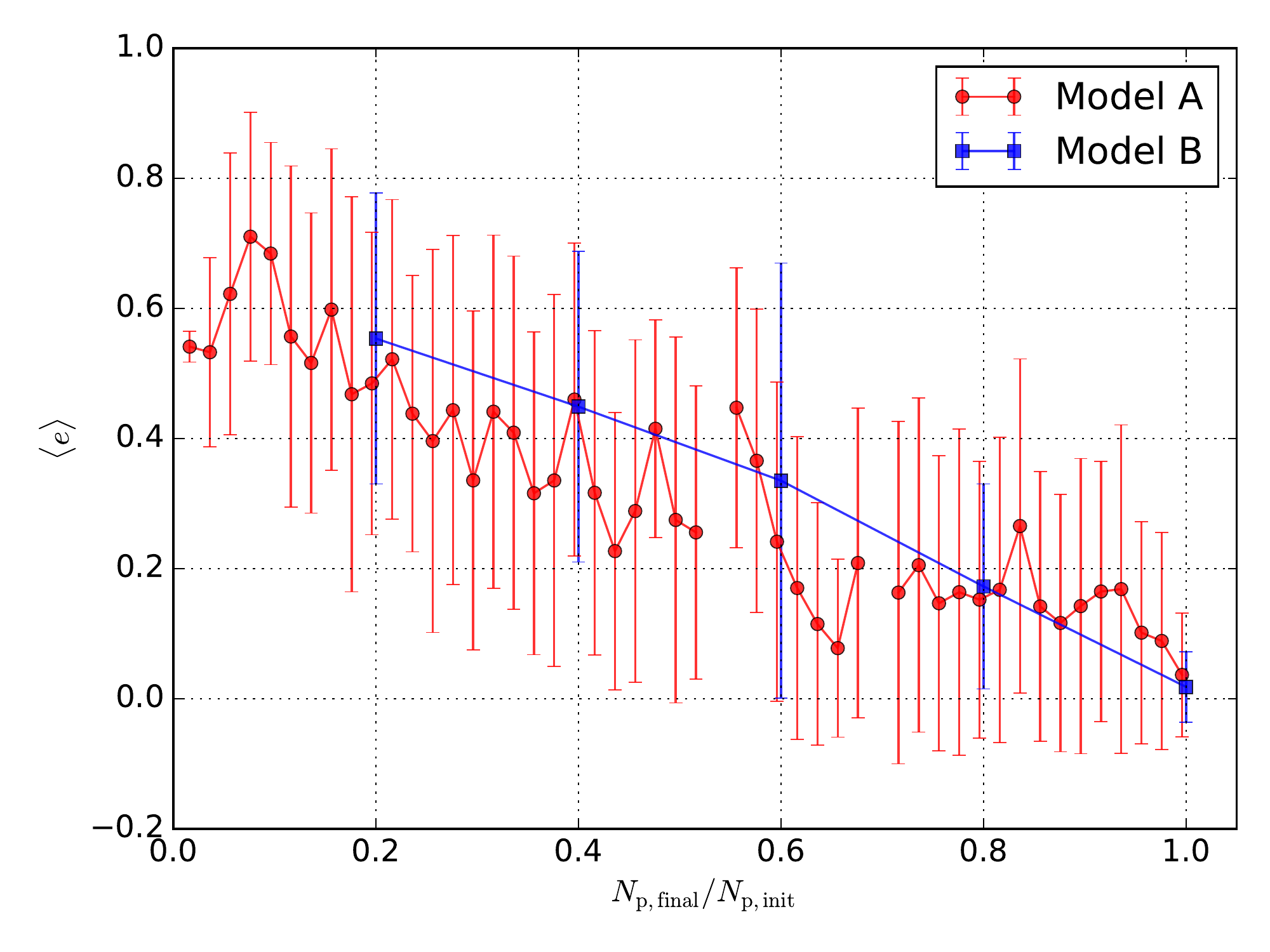}
	\caption{The mean eccentricity and the standard deviation of eccentricities (shown as errorbars) as functions of the fraction of surviving planets ($N_{\rm p, final}/N_{\rm p, init}$). The discontinuities are due to the lack of planetary systems. The climbing trend in the far-left of the plot is due to small-number statistics. Note that the error bars show the magnitudes of the standard deviation of eccentricities, and therefore the negative values in the lower parts of some error bars do not mean negative eccentricities.  }
    \label{fig:ecc_dispersion}
\end{figure}

\begin{figure}
	\centering
	\includegraphics[scale=0.4]{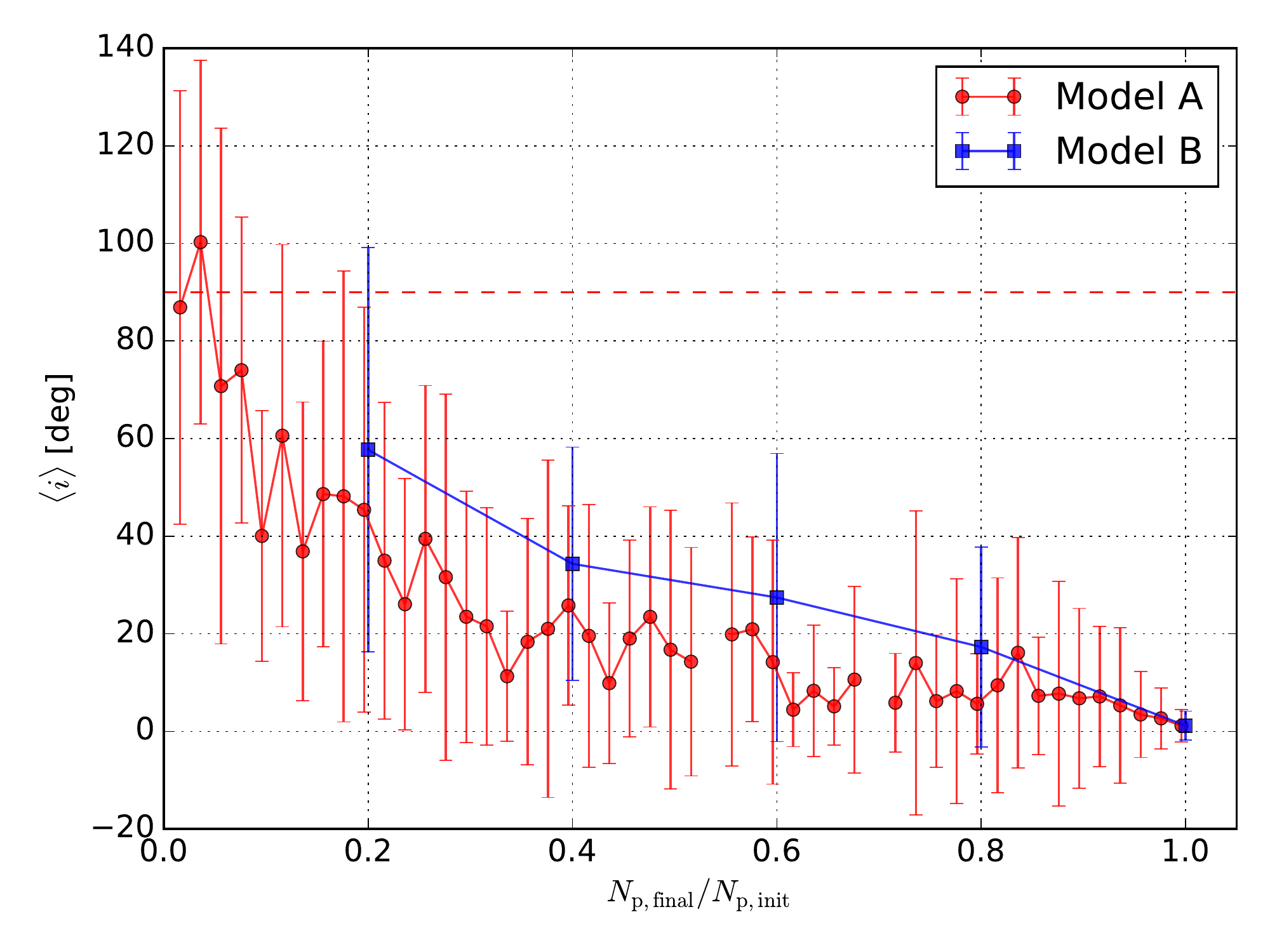}
	\caption{The mean inclinations and the standard deviation of inclinations (shown as errorbars) as functions of the fraction of surviving planets. Notation same as Fig.~\ref{fig:ecc_dispersion}. Note that the error bars show the magnitudes of the standard deviation of inclinations, and therefore the negative values in the lower parts of some error bars do not mean negative inclinations. }
	\label{fig:inc_dispersion}
\end{figure}

For the surviving planets of both Model-A and Model-B systems, the mean orbital eccentricities and the standard deviation of eccentricities are shown in Fig.~\ref{fig:ecc_dispersion}. The figure essentially states that if a planetary system manages to retain $N_{\rm p}$ planets in the star cluster (at the end of the simulation), its mean orbital eccentricity should not be higher than the value given at the central point of the error bar. For instance, in order for all planets in a planetary system to survive, the system must be largely ``untouched''  by stellar encounters such that the mean eccentricity is of the order of $\langle e \rangle \sim 0.05$. Similarly, Fig.~\ref{fig:inc_dispersion} shows the mean inclinations (and their standard deviations) as a function of $N_{\rm p}$. The inclinations are measured with respect to the primordial orbital plane of the planetary systems when the simulations start. For both models, there is an anti-correlation between the dynamical temperature of the system and the multiplicity: hotter systems have a lesser degree of multiplicity, whereas cooler systems have higher degrees of multiplicity. As a consequence of angular momentum exchanges, Model-A systems have slightly smaller $\langle e \rangle$ and $\langle i \rangle$ but larger standard deviation compared to Model-B.

\subsection{Survival Rates as a Function of Time and Initial Semi-major Axes}
\begin{figure}
	\centering
	\includegraphics[scale=0.45]{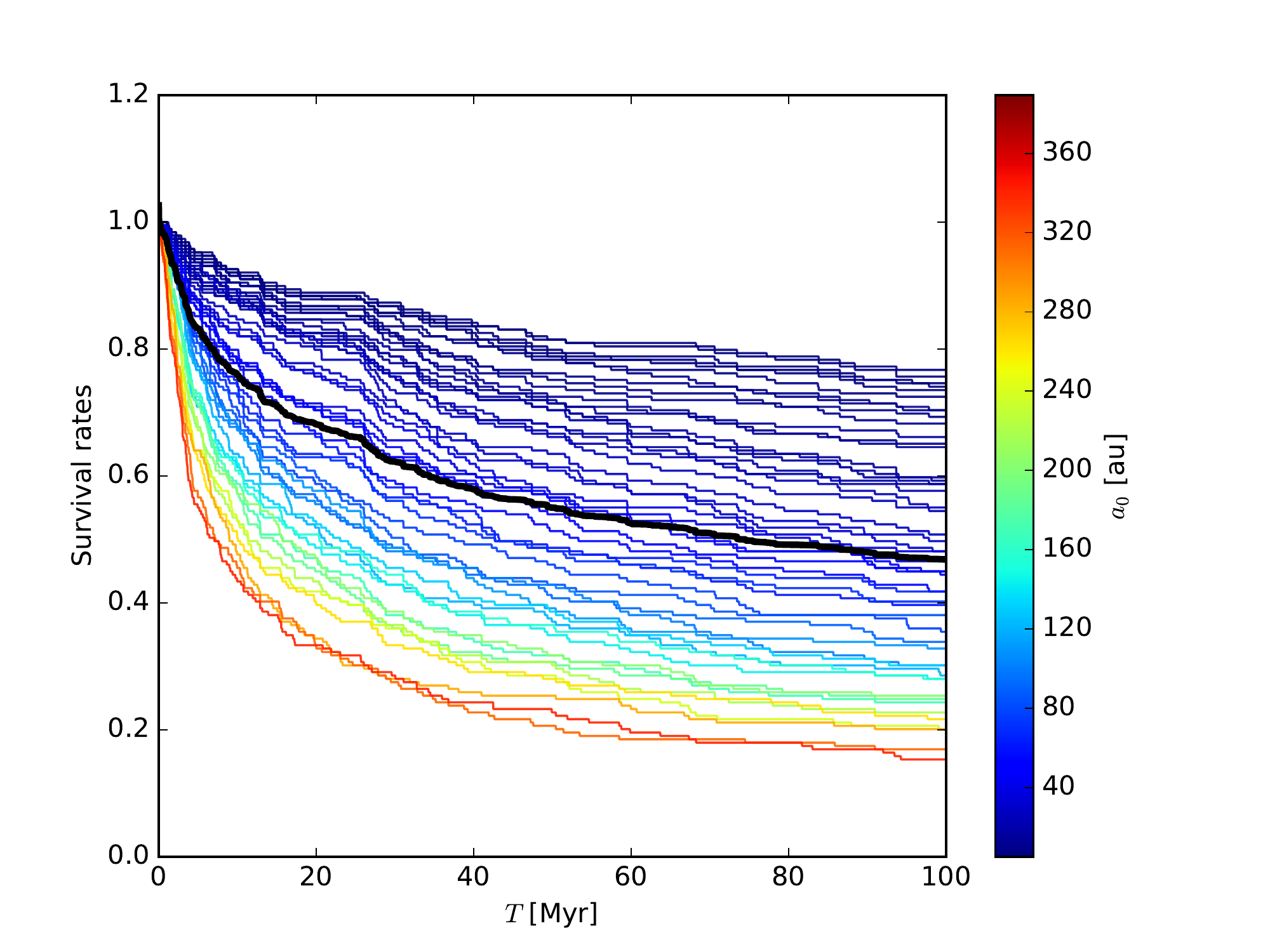}
	\caption{Survival rates of individual planets in YMCs as a function of time (Model-A). The initial semi-major axes of planets $a_0$ are encoded in colors. The thick black curve corresponds to the overall survival rates, which is defined as $N_{\rm p,final}/N_{\rm p, init}$. The uppermost curve generally corresponds to the innermost planet (which, in turn, corresponds to the leftmost part of Fig.~\ref{fig:f_surv_a}), and the lowermost curve generally corresponds to the outermost planet (corresponds to the rightmost part of Fig.~\ref{fig:f_surv_a}).}
	\label{fig:survival_rate_model_a}
\end{figure}

\begin{figure}
	\centering
	\includegraphics[scale=0.4]{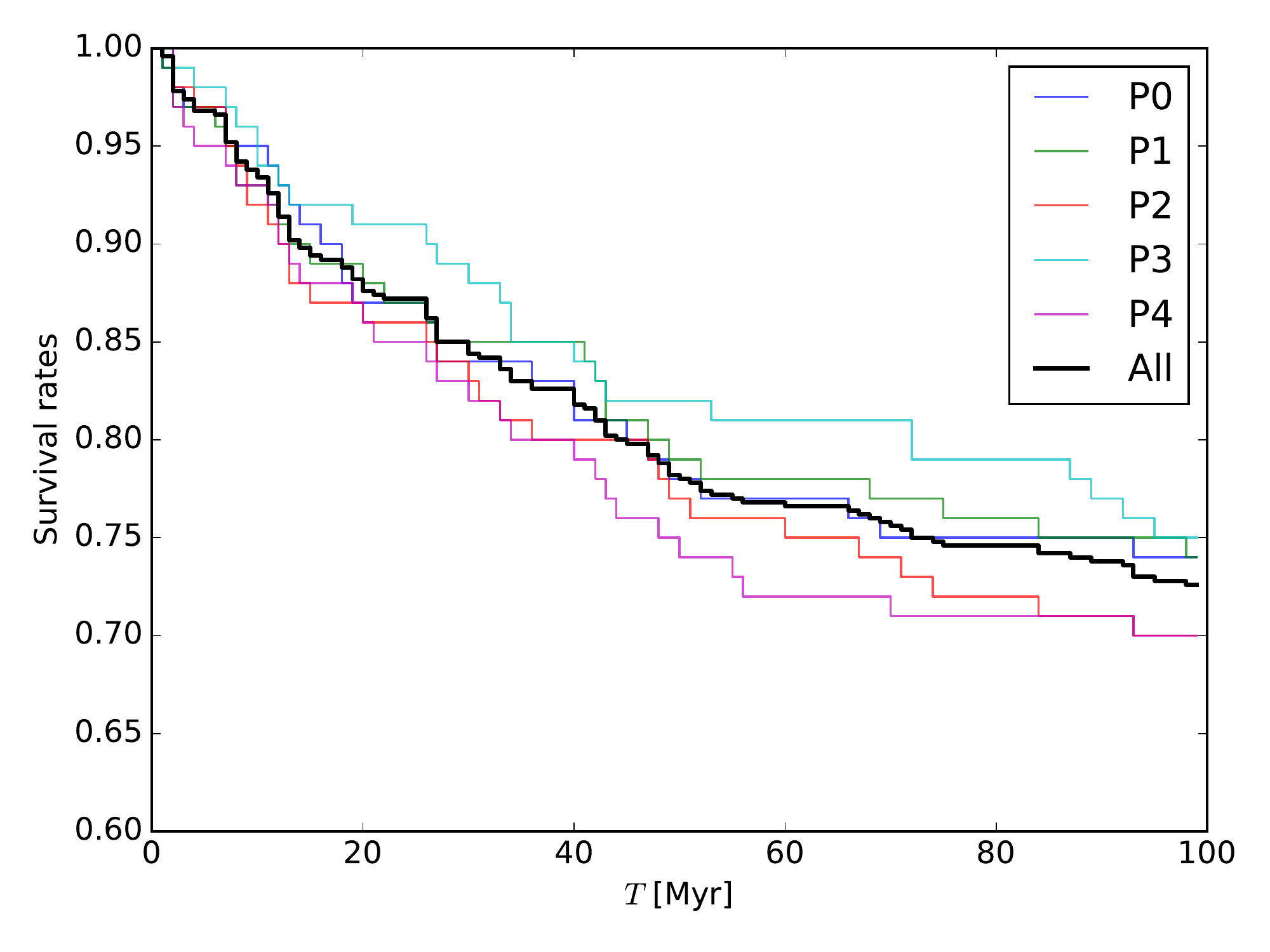}
	\caption{Survival rates of individual planets in YMCs as a function of time (Model-B). P0 is the innermost planet with an initial semi-major axis of 0.5 au, and P5 is the outermost planet in this model with an initial semi-major axis of 6 au.}
	\label{fig:survival_rate_model_b}
\end{figure}

It is clear from the previous subsection that initial semi-major axes matter to the survival rates. Now we can consider the time-dependence. The profiles of survival rates as a function of time for Model-A systems and Model-B systems are plotted in Fig.~\ref{fig:survival_rate_model_a} and Fig.~\ref{fig:survival_rate_model_b}, respectively. The survival rate drops rapidly during the first few Myr (especially those wide planets in Model-A), and then the decline becomes more gradual in due time. Again, the natural selection process provides a feasible explanation for this behavior: a large number of planets are eliminated in the beginning since they are unfit in dense stellar environments, but the more resilient ones prevail. In due time, the ejection rates become increasingly gradual since the surviving planets are more difficult to eject. As such, the remaining planets at the end of the simulations generally reveal the suitable orbital architectures for surviving in dense stellar environments. The planets in Model-A with a wide range of initial semi-major axes generates a series of decay profiles, which can be fitted well with a family of functional form of exponential decay:
\begin{equation}
	f_{\rm surv}(t, a_0) = -0.33 \log_{10}(a_0) \left(1 - e^{-0.0482t} \right) + 1,
	\label{eq:f_surv_t_a}
\end{equation}
where $a_0$ is the initial semi-major axis in au, $t$ is the simulation time in Myr. The fit is valid for a range of semi-major axes from 5 au to 350 au. The coefficient $-0.33$ characterizes the variation of $f_{\rm surv}$ (see also Fig.~\ref{fig:f_surv_a} and Eq.~\ref{eq:f_surv_a0}), where as the power index $-0.0482$ characterizes the ejection rates (see the exponential decay behavior of $f_{\rm surv}$ in Fig.~\ref{fig:survival_rate_model_a}). The $1\sigma$ fitting error of these two aforementioned parameters are $0.00686$ and $0.00226$, respectively. In the limit of $t = 0$, Eq.~\ref{eq:f_surv_t_a} yields a survival rate of $f_{\rm surv} = 1$ regardless of $a_0$; in the limit of $t \rightarrow \infty$, Eq.~\ref{eq:f_surv_t_a} gives an upper estimation of the fraction of planet population that survived from the natural selection process. With a given $t$, $f_{\rm surv}$ is a linear function of $\log_{10}(a_0)$, and at $t = 100$ Myr, Eq.~\ref{eq:f_surv_a0} is recovered. With a certain $a_0$, the survival rates exhibits roughly exponential decay behavior, suggesting a systematic slow-down of ejection rates at large $t$. However, caution should be taken when applying Eq.~\ref{eq:f_surv_t_a} to $t \rightarrow \infty$, because it is certainly possible that strong encounters may occur beyond the time limit of our simulation and lead to the ejection of more planets, and therefore the upper limit provided by Eq.~\ref{eq:f_surv_t_a} simply assumes that the stellar environments remain roughly unchanged beyond the timescale of our simulations. When using Eq.~\ref{eq:f_surv_t_a} to estimate the survival rates as a function of $a_0$ and $t$, the equation overpredicts the survival rates by $\sim 20\%$ for for $t < 15$ Myr, but subsequently converges to an error of $\sim 5\%$ for $t > 20$ Myr.


Due to the planet-planet scattering process in Model-B systems, external perturbations is no longer the sole driver of planet ejections. It is interesting to notice in Fig.~\ref{fig:survival_rate_model_b} that the innermost planet is not necessarily the ones that have the highest chance to survive. This result is in agreement with an earlier study by \cite{2017MNRAS.470.4337C}, where they simulated smaller clusters of $N=2000,8000,32000$ stars.


\subsection{Kinematics of Free-Floating Planets}
\begin{figure}
	\centering
	\includegraphics[scale=0.55]{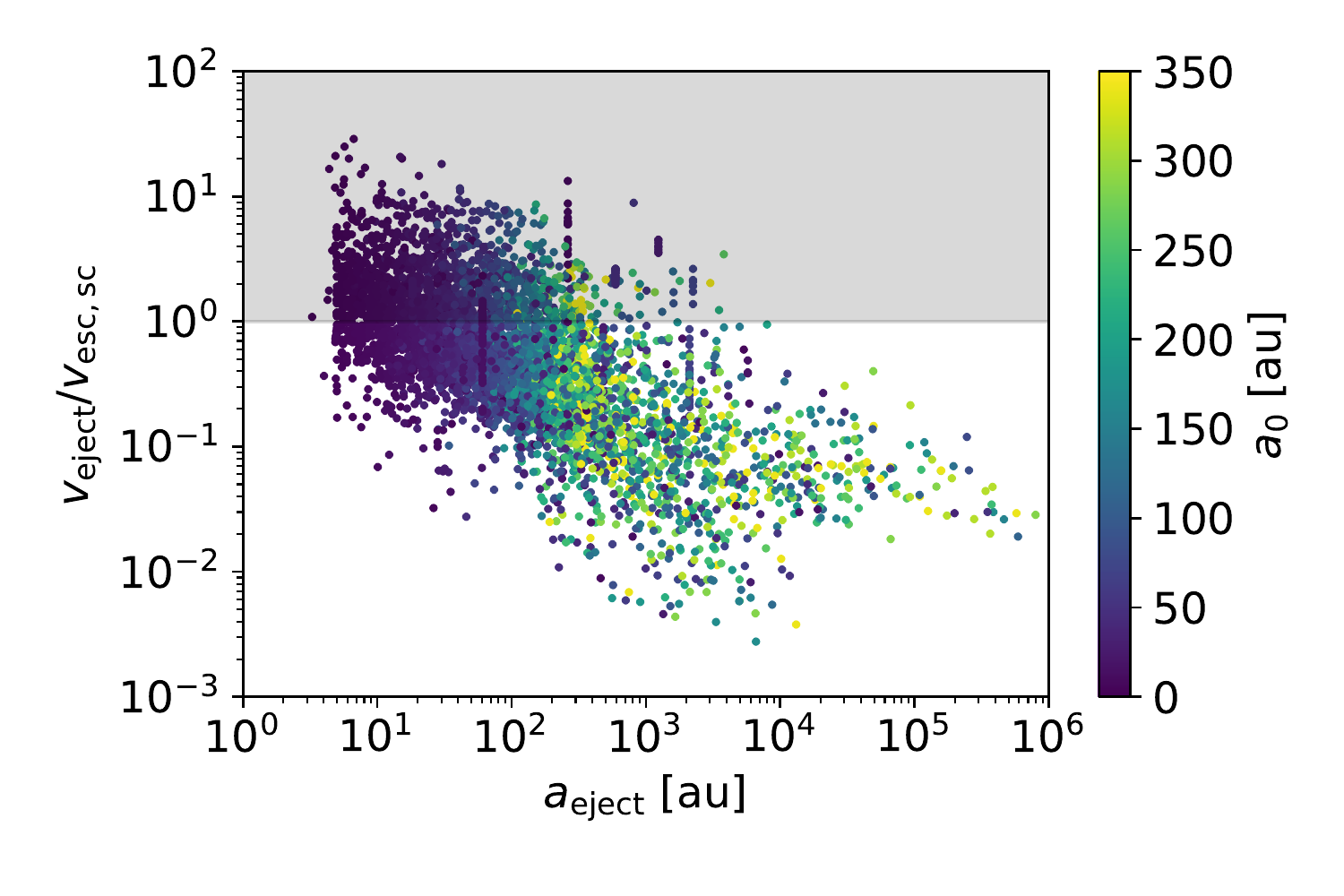}
	\caption{Ejection speed of free-floating planets normalized to the corresponding escape speed of the host cluster (at the location where the ejection takes place). The x-axis is the semi-major axis of the planets prior to their ejections. The color encodes the initial semi-major ($a_{\rm init}$) axis of ejected planets. There are two populations of free-floating planets: in the shaded area, the ejected planet has sufficient velocity to escape the host cluster immediately ($\sim 28.8\%$); below the horizontal line, the ejected planet does not have sufficient velocity to escape the host cluster ($\sim 71.2\%$). }
	\label{fig:v_eject}
\end{figure}

Should a planet be ejected due to the perturbation of stellar encounters, a natural question arises regarding its destination: will the planet remain gravitationally bound to the host cluster, or will it escape? In Fig.~\ref{fig:v_eject}, we plot the ejection speeds of free-floating planets as a function of their semi-major axis immediately prior to the ejection ($a_{\rm eject}$). The initial semi-major axes at $t=0$ are shown in colors. The ejection speeds are scaled to the escape velocity of the Plummer sphere at the point of ejection \citep{2003gmbp.book.....H}:
\begin{equation}
	v_{\rm esc,sc}(r) = \left( \frac{2 G M}{r} \right)^{1/2} \left( 1 + \frac{r_{\rm hm}^2}{r^2} \right)^{-3/4},
	\label{eq:v_esc_sc}
\end{equation}
where $r$ is the distance from the planet ejection location to the host cluster center, $r_{\rm hm}$ is the half-mass radius of the cluster, $G$ is gravitational constant, and $M$ is the total mass of the cluster.

It is evident from Fig.~\ref{fig:v_eject} that high ejection velocities are produced from planets with smaller $a_{\rm eject}$, where the orbital velocities are higher. There are two ejection velocity regimes: within the shaded area, the ejected planet has sufficient velocity to escape from the cluster, and therefore once they are ejected from their planetary systems, they are also instantaneously become unbound to the cluster potential; below the line, the planets only have enough velocity to escape from their planetary systems, but they do not have sufficient velocity to escape from the host cluster, and therefore they will be free-floating planets inside the cluster. About 1/3 ($\sim 28.8\%$) of ejected planets have sufficient speeds to escape from the host cluster immediately. This population, namely ``prompt ejectors'', mostly originate from the inner planetary systems with small semi-major axes. More than 2/3 of ejected planets ($\sim 71.2\%$) are the so-called intra-cluster free-floating planets, which, due to their low masses, may be subsequently expelled from the host cluster through the mass-segregation process that takes place at a timescale of $\sim 200$~Myr for the YMC model used in our simulations \citep{1971ApJ...166..483S,2002ApJ...565.1251H,2006ApJ...641..319P,2010ARA&A..48..431P}. 

In a recent study, \cite{2015MNRAS.449.3543W} simulate the dynamics of free-floating planets in an $N \sim 2000$ star cluster using the GPU-accelerated direct $N$-body package \texttt{NBODY6} \citep{2003gnbs.book.....A,2012MNRAS.424..545N}. They observe that the planet-to-star ratio drops rapidly by half in the first few Myr, and then followed by a steady and linear decline in the next 1.6 Gyr until the population of free-floating planets depletes. The authors argue that the rapid decay in the first few Myr is a consequence of direct ejections, whereas the steady decline is a result of the mass segregation process. Since the cluster tends to establish an equal-partition of energy, low-mass objects will, therefore, obtain high velocities and eventually escape host cluster. We expect a large population of free-floating planets in any galaxy. Nevertheless, it is important to note that the free-floating planet population has already existed at the beginning of the simulations in \cite{2015MNRAS.449.3543W}, where they sample the initial positions and velocities of the planets from the same distribution os that of the stars. In a new study by \cite{2019arXiv190204652V}, the free-floating planet population is generated in a more self-consistent way, as they model all planets to be initially bound to host stars, and that free-floating planets are only created upon ejections. Their simulations with a super-virial ($Q=0.6$) and fractal ($F=1.26$) cluster yield the prompt ejector population and the delayed ejector population as well. More interestingly, they are able to simulate the recapturing of free-floating planets. They conclude that a small fraction ($\sim 1.5\%$) of free-floating planets are subsequently recaptured by another star. Since this fraction is low, and that the recaptured planet is typically on highly eccentric and inclined wide orbits (which are prone to perturbations) \citep{2015MNRAS.453.3157J}, we do not think that recapturing is an effective mechanism to help YMCs to retain planets.

Interestingly, we also observe that many planets (especially those with large $a_{\rm init}$) have undergone substantial outward migrations. The energy injected by stellar flybys has elevated the semi-major axes of outer planets (and even a small number of inner planets), making the planetary systems increasingly ``fluffy'' as a function of time. Therefore, dense stellar environments can lead to an outflow of planets/massless-particles. Planets and low-mass particles can either be ejected \emph{in situ} at roughly their initial semi-major axes, or undergoes outward migration and subsequently be ejected at larger semi-major axes, and thereby generate free-floating objects such as 1I/'Oumuamua \citep[e.g.,][Torres et al. 2019 (submitted)]{2018MNRAS.479L..17P,2019arXiv190102465H}.

In the planetary systems where planet-planet scattering is important (e.g. Model-B systems), the ejection of planets is not solely caused by external perturbations. A series of moderate or weak encounters may gradually increase the angular momentum deficits \citep[][]{1997A&A...317L..75L} of an externally perturbed planetary system, and then a planet may get ejected by another planet during a close encounter event long after the stellar encounter. Indeed, the planet ejection efficiency is roughly doubled in this ``delayed ejection'' scenario \cite{2017MNRAS.470.4337C}.

\subsection{The Effects of Binary Perturbers}
\begin{figure}
	\centering
	\includegraphics[scale=0.55]{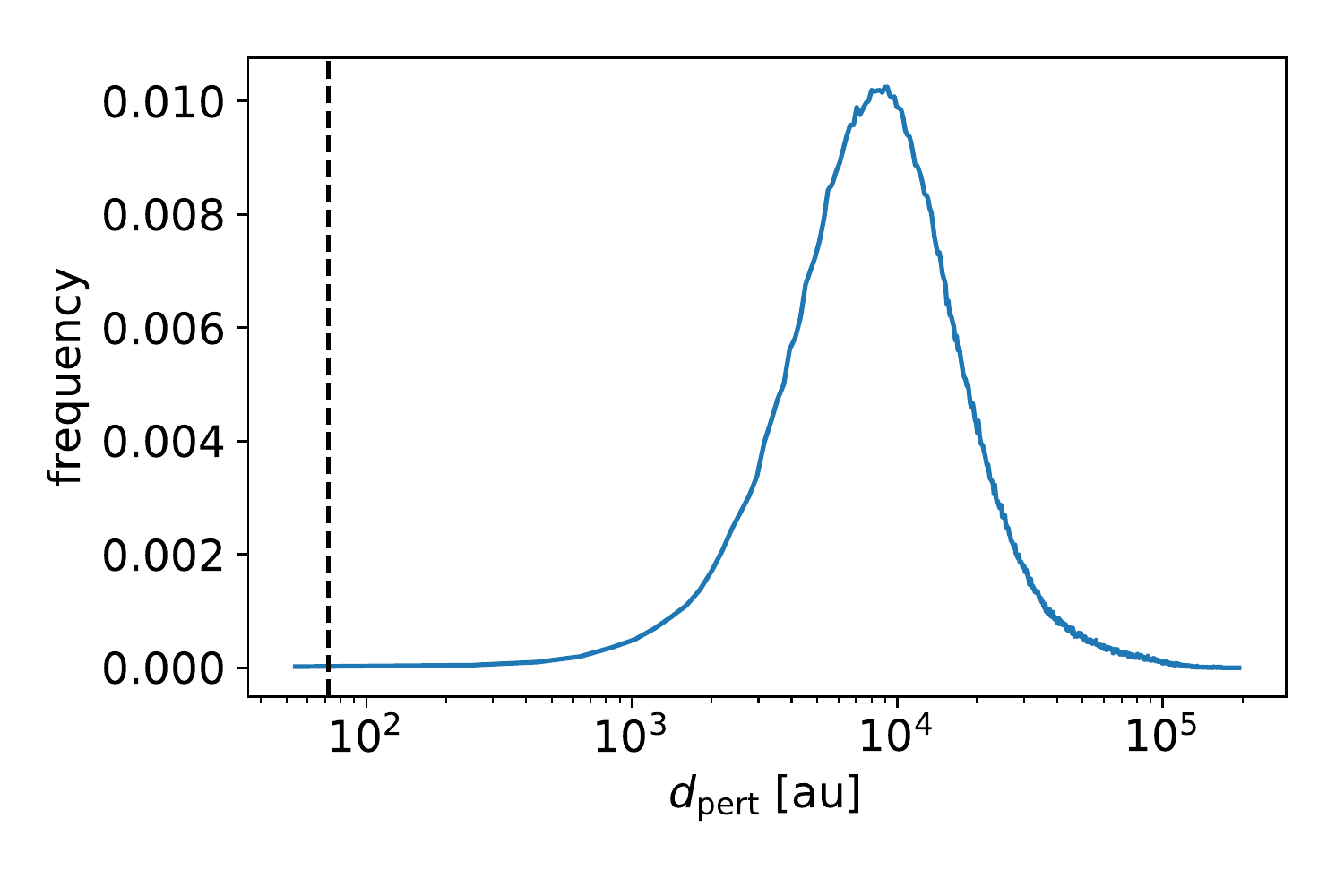}
	\caption{Distribution of the distances of the closest perturber. The vertical line indicates the maximum initial semi-major axis of the primordial binary population.}
	\label{fig:pert_dist}
\end{figure}

Our host cluster consists of $10\%$ of primordial binaries with the initial semi-major axes $a_{\rm bin}$ ranging from [0.007, 72] au ($\log(a_{\rm bin})$ has a uniform distribution, see Section~\ref{sec:modeling_ic}) , which is by far smaller than the median distance of the closest perturber (of the order of $10^4$~au), as shown in Fig.~\ref{fig:pert_dist}. With most binary systems being hard binaries, they behave dynamically as if single perturbers \citep{2009PASJ...61..721T} that double their masses (since the mass ratio is 1). In the rare event where a binary is sufficiently close to a planetary system such that its quadruple moment becomes important to the dynamics of the planetary system, a recent Monte-Carlo scattering experiment by \cite{2015MNRAS.448..344L} has shown that the cross section for eccentricity increase would be a factor of $2-3.6$ compared to single-star encounters.

\subsection{The Uncertainty Due to the Rareness of Strong Encounter Events}
Two-body relaxation in star cluster is essentially a Coulomb scattering process \cite{1994ApJ...424..292O}, the cumulative effects of rare but strong encounters on exciting the planetary orbital eccentricities is comparable to the commutative effects frequent but weak encounters. The rareness of strong encounters may introduce a statistical uncertainty should we initialize our host cluster with a different random seed.


Recently, \cite{2018arXiv181108598F} use semi-analytical formulation to estimate the survival rates of single-planet systems in the Pleiades, Hyades, and Praesepe open clusters. They point out that, for an encounter event, clost-in planets within 1~au will have less than $1.5\%$ probability to be ejected, and for planets within 10~au the probability of ejection due to this encounter will be at most $7\%$. For single-planet systems, close-in planets can only be ejected by strong encounters. This, in turns, can be translated into the frequency of strong encounters being just a few percents. As such, we do not expect that the statistical noise in the strong encounter regime to change our main result, since the robustness of our results is mainly supported by the prevalence of weak-to-moderate encounter events.

\section{Discussions}
\label{sec:discussions}
\subsection{Planet Formation in Dense Cluster Environments}
The planet formation process, which takes place in protoplanetary discs, may be complicated by the high stellar density and intense radiation fields in YMCs. While our simulations cover a wide range of initial semi-major axes from 0.5 au to 350 au, the perspectives of planet formation in extreme environments is still under active debates. There are several dedicated studies on the effects of stellar environments to the protoplanetary discs. For example, a number of authors suggest that the protoplanetary disc can be truncated, or even disrupted by close stellar flybys \citep[e.g.,][]{2012ApJ...756..123O,2016MNRAS.457..313P,2016ApJ...828...48V,2017A&A...604A..91W,2018MNRAS.477.5191R,2018ApJ...868....1V}, although the viscous evolution of the disc may be helpful in damping the eccentricity and eventually establishing a new equilibrium \citep{2019MNRAS.482..732C}. On the other hand, the initial mass function dictates that there will be a small fraction of O/B stars that emit energetic FUV photons. Due to the proximity, these FUV photos can photoevaporate the disc \citep{2013ApJ...774....9A,2018MNRAS.481..452H,2018MNRAS.478.2700W}, which may prematurely halt the core accretion process and the disc-driven process.

\subsection{Enhanced planet ejection efficiency due to the structural evolution of the host cluster}
It is worth mentioning that the structure of the YMC is evolving quickly due to the gas expulsion \citep[e.g.,][]{2006MNRAS.373..752G, 2008MNRAS.384.1231B, 2016A&A...587A..53K}. Once the intra-cluster gas is gone, the cluster becomes supervivial temporarily, and the subsequent reestablishment of virial equilibrium may lead to a temporary surge of planet ejection rates. In a recent study, \cite{2015MNRAS.453.2759Z} carry out direct $N$-body simulations of clusters with different initial morphologies and initial virial states. They conclude that non-equilibrium conditions in host clusters can indeed boost the ejection rate, and that for a cluster of $N=1000$ stars, a virial equilibrium is reestablished at an energy equal-partition timescale of 5~Myr, after which the ejection rates becomes steady again. 

\subsection{The recapturing of free-floating planets}
Given the high stellar density, it is possible that an ejected planet can be recaptured by another star. \cite{2012ApJ...750...83P} suggest that if recapturing does happen, planets are likely to be captured into wide orbits with the new semi-major axes of the order of $10^2 - 10^6$ au. In this sense, the probability for the recaptured planet to stay bound to its new planetary system depends on how long the new planetary system stays in the host YMC. If the new planetary system stays in the YMC for an extensive period of time, the chances of getting re-ejection would be high. However, if the new planetary system has already escaped from the host YMC, then the recaptured planet may be able to stay bound over secular timescales. In a related study, \cite{2015MNRAS.453.3157J} suggest that the dwarf planet Sedna in the Solar System, as we as many Sednitos, may be a result of recapturing. In a related study, \cite{2011MNRAS.411..859M} suggest another scenario where up to a few percents of low-mass intruders may themselves be captured and become bound to the host star, which can potentially lead to drastic changes in the orbital properties of the planets orbiting the stars.

\subsection{The planetary mass-metallicity correlation}
According to the planet-metallicity correction \citep{2005ApJ...622.1102F}, the GC environment, which is typically metal-poor, is not ideal for the formation of Jovian planets. However, the formation of terrestrial planets is less affected by the metallicity \citep{2015AJ....149...14W}. In this sense, we suspect that the lack of planet detection in dense GCs such as 47 Tuc \citep{2000ApJ...545L..47G,2017AJ....153..187M} is in fact due to the low frequency of the more-easily detectable Jovian planets. We speculate that short-period terrestrial planets can be found in dense GCs with next-generation surveys. 

\subsection{The survivability of short-period planets}
While most short-period planets ($a_0 \leq 1$) are stable over the 100 Myr timescale of our simulations, their long-term survivability can be complicated by multiple physical processes. For example, a planet may be engulfed when its host star evolved to the post-main sequence stage \citep[e.g.,][]{2001Natur.411..163I, 2006Natur.442..543M, 2010MNRAS.408..631N}. In certain cases, the existence of an inclined stellar companion or outer planet may trigger the Lidov-Kozai oscillation, pushing a close-in planet to a highly eccentric orbit. In a multi-planet system, planet-planet scattering may lead to ejections \citep[e.g.,][]{2017MNRAS.470.1750I} and/or mutually inclined orbits \citep[e.g.,][]{2016ApJ...822...54D,2019arXiv190402290X}. Nevertheless, it is evident that these complications will only affect the survivability of a fraction of planets. For example, if a planet orbits a low-mass star (which is highly likely given the power-law of the initial mass function), one could expect that the host star will remain in the main sequence longer than the typically age of the a GC; if the planet is not in a circumbinary binary system, the probability for two single stars to dynamically form a binary system and subsequently trigger the Lidov-Kozai oscillation would be low. Therefore, we expect that most of the surviving short-period planets remain bound to their host stars over Gyr timescales. 

\subsection{Extrapolating the results to star clusters of different sizes}
\begin{figure}
	\center
	\includegraphics[scale=0.5]{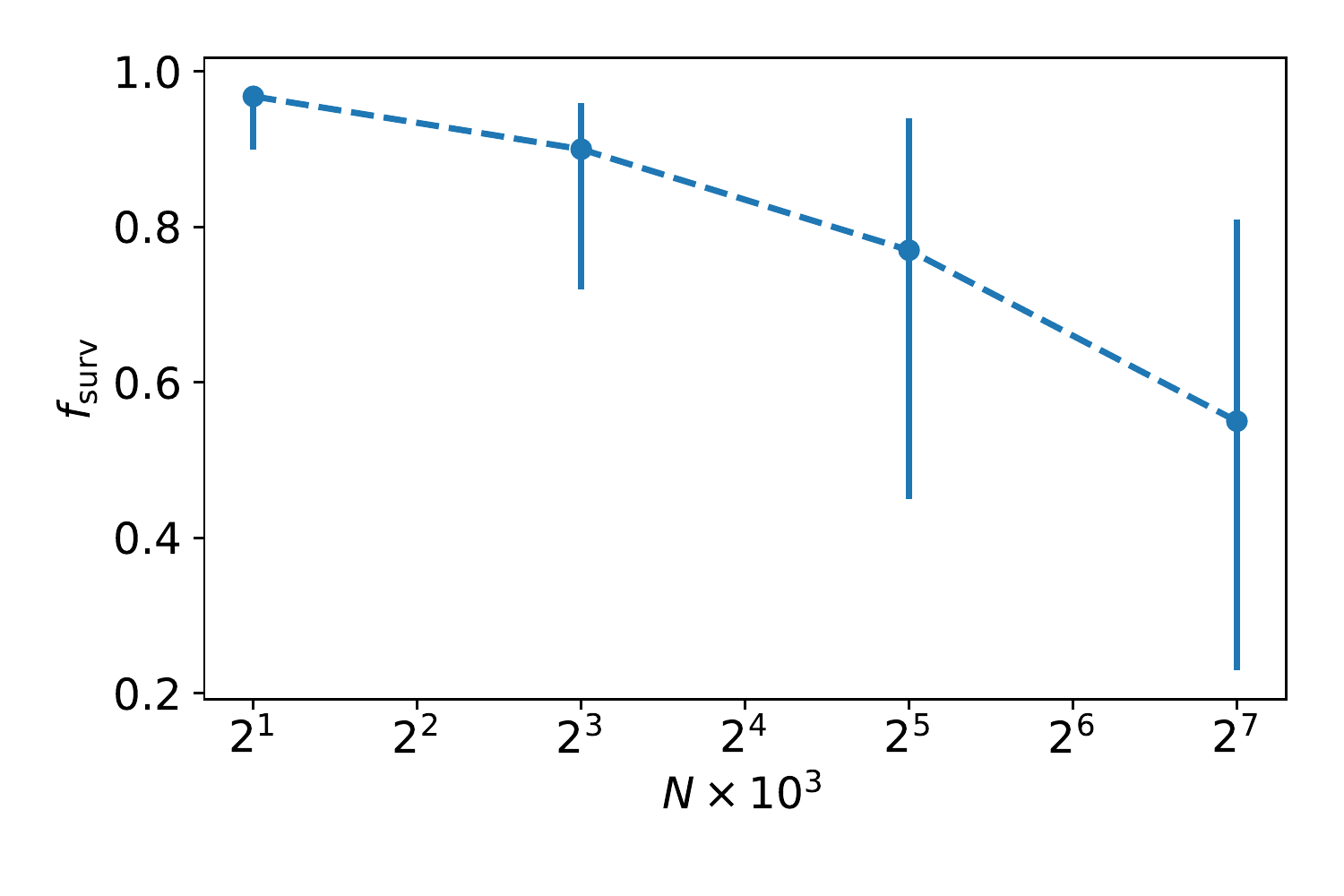}
	\caption{Overall survival rates of planets in star clusters of different $N$, obtained through the compilation of the results from Cai et al. 2017 (Model IV systems, $N = 2{\rm k}, 8{\rm k}, 32{\rm k}$ and a fixed virial radius of 1 parsec) and this study $N=128{\rm k}, r_{\rm vir} = 1.74$ parsec at $t = 50{\rm Myr}$. The mean stellar density of these four cluster models is roughly proportional to $N$. Due to the slightly different initial conditions used in Cai et al. 2017 Model IV and the Model A of this paper, we obtain the survivability of the $N=128{\rm k}$ model only from a subset of planets that have the initial conditions directly comparable to Cai et al. 2017. The central dots of the error bars is the overall survival rates, and the upper/lower limits of the errors bars correspond to the survival rates of planets with $a_0 = 5$ au and $a_0 = 200$ au, respectively.}
	\label{fig:f_surv_N}
\end{figure}

This study focuses on the dynamics of exoplanet in a YMC of $N = 128{\rm k}$ and $r_{\rm vir} = 1.74$ parsec, comparable to the Westerlund-1 cluster. In an earlier study, \cite{2017MNRAS.470.4337C} carried out a parameter-space survey of the survivability of planets in star cluster of different $N$. Although they only evolve their simulations up to 50 Myr and the results are based on slightly different initial conditions (semi-major axis range from 5.2 to 206 au), their results allow us to obtain a scaling relationship of the planet survivability as a function of $N$, as shown in Fig.~\ref{fig:f_surv_N}. Extrapolating from this trend, the overall survival rates for an $N = 512 {\rm k}$ cluster would be below 0.3, if planetary systems formed in such a dense cluster is still similar to the orbital architectures that we use in this study. Moreover, it is evident from the figure that the range of $f_{\rm surv}$ (denoted by the length of the error bars) is larger with bigger clusters, which is indicative that the diversity of exoplanets increases with denser and larger clusters.  If we generalize Eq.~\ref{eq:f_surv_t_a} as
 \begin{equation}
	f_{\rm surv}(t, a_0) = -\alpha \log_{10}(a_0) \left(1 - e^{-\beta t} \right) + 1,
	\label{eq:f_surv_t_a_generalized}
\end{equation}
Where $\alpha > 0$ and $\beta > 0$, then one could expect that these two coefficients both depend on $N$. As $N$ increases, the spreading of $f_{\rm surv}$ increases, and so $\alpha$ (essentially defined as $\alpha \equiv d f_{\rm surv} / d \log_{10}(a_0)$, see Fig.~\ref{fig:f_surv_a}) increases. At the same time, it is likely that ejection is more efficient in a dense cluster, and so $\beta$, which characterizes the ejection efficiency, also increases.   

\subsection{Limitations of the results in constraining the exoplanet population in globular clusters}
Cautions should be taken when interpolating the results of this work to the properties of exoplanets in GCs. First, the evolution timescale of observed GCs (multi-Gyr) is much longer than the timescale of our 100 Myr simulations. During the vast timescale of GC evolution, many physical processes, such as stellar evolution, dynamical perturbation, mass segregation of the host cluster, and the stat-planet tidal evolution can all complicate the planetary evolution. Second, the ancient YMCs that eventually evolved into the present-day GCs may have different properties (in particular, metallicity) comparing to the present-day YMCs. Last, even though recent surveys \citep[e.g.,][]{2019MNRAS.482.5138B} of GCs provide us detailed properties of the present-day GC population, we are not yet able to fully reconstruct the initial conditions of these GCs back to a few Gyr ago. Therefore, with the lack of knowledge about the initial conditions and the constrains of computing time, our results of YMC planet survivability can only be interpreted as an upper limit of the GC planet survivability.


\section{Conclusions}
\label{sec:conclusions}
As of July 2019, with more than 4,000 confirmed exoplanets, there is only one exoplanet detected in the dense globular cluster (GC) Messier 4. Motivated by this, we launch direct $N$-body simulations using the \texttt{AMUSE} framework to investigate the role of dynamical perturbations in shaping the planetary systems in dense GC environments. Our simulations start from a YMC-like environment, since it is plausible that YMCs are the progenitors of GCs. We address two fundamental problems regarding whether YMCs can harbor planets and whether GCs can inherit planets from YMCs if there are indeed evolved from YMCs. Our main conclusions are summarized as below:
\begin{itemize}
	\item The dense stellar environments in YMCs do have profound effects in shaping planetary systems. Nascent planetary systems are forced into a natural selection process, and that only the most robust orbital architectures can survive. Dense stellar environments favor planets with short orbital periods, and the survival probability as a function of the initial semi-major axis $a_0$ (in au) at $t$ (Myr) can be estimated with $f_{\rm surv}(t, a_0) = \left( 0.33 \times 10^{-0.02t} - 0.33 \right) \log_{10}(a_0) + 1$. On average, the survival rates for planets with semi-major axes larger than 20 au is lower than $50\%$ after the 100 Myr evolution. With a fixed $f_{\rm surv}=0.5$, the mean stellar density (in the units of $M_{\odot}/{\rm pc}^3$) roughly scales with the $a_0$ as $\langle \rho (a_0) \rangle = 3500 \exp(-0.007a_0)$, indicating that only planets with very tight orbits can survive in the dense regions of the host cluster. 
    \item Planetary systems with high multiplicity can only be found in the outskirts of YMCs or in the field. In contrast, the dense cluster center produces ``hot'' planetary systems with low-degree of multiplicity and high mean eccentricities/inclinations. The host cluster shapes its planetary system through a natural selection process in that only the most suitable orbital architectures survive, and the survivability of planets as a function of time can be well approximated with an exponential decay function.
    \item Weak stellar encounters produce ``cluster wanderers'', whereas strong encounters produce ``cluster escapers''. Among those free-floating planets (FFPs) produced by dynamical ejections, $\sim 28.8\%$ of them have sufficient velocities to escape from the host clusters. These cluster escapers are mostly originated from the inner regions of the original planetary systems where the orbital speeds are high. Ejecting these short-period planets requires strong encounters, which are rare events. The prevalence of medium-to-weak encounters is only able to eject wide planets, whose low orbital velocities will cause them to be confined in the host clusters. However, ``cluster wanderers'' may be gradually expelled by the subsequent mass-segregation process in the host cluster.
    \item Free-floating planets (FFPs) in the Galactic field can be generated through four channels: (1) direct ejection from the inner regions of the original planetary systems through strong encounters; (2) ejected from the original planetary systems while remaining bound to the cluster, until being expelled by the mass-segregation process; (3) ejected from the original planetary systems and remain bound to the cluster until the dissolution of the host clusters; (4) ejected directly from an isolated planetary systems in the Galactic field, whose angular momentum deficit was stirred up by stellar encounters when the planetary systems were still in the host cluster. 
\end{itemize}

If the hypothesis that YMCs being the progenitors of GCs is true, our results suggest that wide-orbit planets and free-floating planets are unlikely to be found in the present-day GCs, but the existence of short-period ($a \sim 0.1$ au) terrestrial planets in dense GCs cannot be ruled out. We speculate that the lack of planet detection in dense GCs is actually due to the lack of Jovian planets in metal-poor environments, and therefore short-period terrestrial planets can still be found in dense GCs with next-generation surveys. However, a full investigation of planets evolution in GCs across multi-Gyr timescale should be addressed in subsequent studies.

\section*{Acknowledgements}
We thank the anonymous referee for their comments that helped to improve the manuscript considerably. We thank Michiko Fujii, Elena Sellentin, and Ignas Snellen for insightful discussions. This project is supported by SURFsara (Dutch national supercomputing center) through the SOIL (SURFsara Open Innovation Lab) initiative and the EU Horizon 2020 project COMPAT (grant agreement No. 671564). The simulations are carried out on the supercomputer \texttt{Cartesius} and the LGM-II GPU machine (NWO grant \#621.016.701). M.B.N.K. acknowledges support from the National Natural Science Foundation of China (grant 11573004). This research was supported by the Research Development Fund (grant RDF-16-01-16) of Xi'an Jiaotong-Liverpool University (XJTLU). We acknowledge the support of the DFG priority program SPP 1992 ``Exploring the Diversity of Extrasolar Planets (Sp 345/20-1)''. This project makes use of the SAO/NASA Astrophysics Data system and the \url{exoplanet.eu} database.




\bibliographystyle{mnras}
\bibliography{refs} 



\appendix


\bsp	
\label{lastpage}
\end{document}